\documentclass{eptcs}

\title{Coroutining Folds with Hyperfunctions}
\author{J. Launchbury, S. Krstic, and T. E. Sauerwein }

\def\titlerunning{Coroutining Folds with Hyperfunctions}
\def\authorrunning{J. Launchbury, S. Krstic, and T. E. Sauerwein}
\newtheorem{theorem}{Theorem}

\begin{document}
\maketitle

\begin{abstract}
Fold functions are a general mechanism for computing over recursive data structures. First-order
folds compute results bottom-up. With higher-order
folds, computations that inherit attributes from above can also be expressed. In this paper,
we explore folds over a form of recursive higher-order function, called hyperfunctions, and
show that hyperfunctions allow fold computations to coroutine across data structures,
as well as compute bottom up and top down. 
We use the compiler technique of foldr-build as an exemplar to show
how hyperfunctions can be used.
\end{abstract}

\section{Dedication}

It is a privilege to submit a paper for the Festschrift symposium held to honor Dave Schmidt's
lifetime of contributions on the occasion of his 60th birthday. Many years ago, as a fresh PhD student, 
Dave's excellent book on Denotational Semantics \cite{Sch86} opened my eyes to rich possibilities
of building functions over functions---recursively!---and our continued interactions over the years were always insightful. So it seemed appropriate to offer a paper whose foundations rely on the same mathematical models that Dave so ably expounded all those years ago, constructing recursive function spaces in a way that is not possible in traditional set-theoretic models. The paper is a revision of an earlier (unpublished) paper \cite{LKS00} whose ideas deserved to see the light of day. -- John Launchbury, 2013.

\section{Introduction}

Folds have been popular for a long time. At the one end of spectrum, they are presented in early classes introducing functional programming. At the other end, they form the foundation of Google's famous world-scale map-reduce computational engine. In this paper, we play with the idea of folds. We use them to introduce and explore a category of coroutining functions, which we style \emph{hyperfunctions}. We do so by tracing the story of a fascinating technique in code fusion, and show how hyperfunctions are able to open up apparently closed doors. Thus, while our story will be about fusion, our narrative purpose is actually to explore hyperfunctions. We will use the Haskell
programming language as our setting.

Code fusion---automatic removal of intermediate computational structure---holds 
a promise of providing the best of two worlds:
programming with the structures enables concise and modular solution to problems;
and the removal of the structures provides efficient run-time implementations \cite{Wad90}.
One particularly enthralling technique is called the \emph{foldr-build} rule
\cite{GLPJ93}. The rule exploits a convergence of three programming
aspects---structured iteration, function abstraction, and parametricity---to
achieve intermediate structure removal in a single transformation step. The transformation
was so effective that it was used for many years within the popular Glasgow Haskell compiler (GHC).
However, a significant shortcoming of the technique is that it has not been clear how to extend
it to fuse \verb@zip@ \cite{LS95}. In this paper, we will introduce hyperfunctions,
lift the foldr-build technique over hyperfunctions, and
show how this enables both branches of zip can be fused concurrently. Thus, even though
the foldr-build approach has been eclipsed in recent years by stream fusion
techniques (which are able to handle zip without problem) \cite{CLS07}, it
is a intriguing demonstration of the power of hyperfunctions that they are able to overcome
a previous shortcoming in what was a significant method for many years. Furthermore, given the
prevalence of fold-like computational structures, folds over hyperfunctions may find other
applications in due course.

\section{Original Foldr-Build}

The key goal of foldr-build is to achieve fusion in one step. It is able to remove 
intermediate computational data structures
without the need for search or extra analysis that are present in many other
techniques. While foldr-build applies to other data structures, it is used
most extensively with lists. We follow this trend, and for the rest of the
paper will focus on lists alone.

The foldr-build idea has four key components: 
\begin{enumerate}
\item List producing functions are
abstracted with respect to cons and nil: for example,
the list \verb@[1,2,3]@ is represented by the function
\verb@\c n -> c 1 (c 2 (c 3 n))@.  
\item List are reconstructed with a
fixed function \verb@build@, defined by \verb@build g = g (:) []@, where the 
\verb@g@ argument is an abstracted list such as written in Step 1.
\item Parametric polymorphism is used
to be sure that abstraction has been complete; this is expressed by requiring \verb@build@ to have
type \verb@build::(forall b . (a->b->b)->b->b)->[a]@, which is an example of rank-2 polymorphism; and 
\item List consumers are defined
using \verb@foldr@. 

\end{enumerate}
An example definition that follows these principles is:
\begin{verbatim}
   map f xs = build (\c n -> foldr (c . f) n xs)
\end{verbatim}
To understand this definition, consider mapping a function \verb@f@ down the list \verb@[1,2,3]@. 
Being explicit about the list constructors, the result is
\verb@(:) (f 1) ((:) (f 2) ((:) (f 3) []))@, or alternatively,
\verb@((:) . f) 1 (((:) . f) 2 (((:) . f) 3 []))@.
The effect of \verb@foldr@ is to replace the conses and
nil of its list argument with the function arguments provided. In this case, the original
conses get replaced with \verb@((:) . f)@, or as we are abstracting over the conses and nils, with,
\verb@(c . f)@.

The foldr-build
theorem below asserts that if a list producer has been properly abstracted with respect
to its conses and nil, then the effect of \verb@foldr@ on the list can be achieved simply
by function application. This is expressed as follows.

\begin{theorem}[Foldr-build]
If \verb@g@ has the polymorphic type \verb@g::forall b.(a->b->b) -> b -> b@ then for all
\verb@k@ and \verb@z@ it is the case that {} \verb@foldr k z (build g) = g k z@.
\end{theorem}
The proof follows easily from the parametricity theorem implied by the
type of \verb@g@ \cite{Wad89,GLPJ93,correctness}. In effect, the polymorphism of \verb@g@ ensures that \verb@g@ behaves uniformly
for every possible substation of its arguments, and that there are no exceptional cases lurking in the
code of \verb@g@ that may behave differently in different settings\footnote{Sadly, 
the presence of pseudo-polymorphic {\tt seq} in Haskell actually causes the theorem to fail, because the resulting parametricity theorem is no longer quite strong enough, even though there
is a general form of parametricity that works for seq \cite{seq-correctness}. However, in practice a simple static check was enough to ensure that the bad situation did not arise.}.

To see the power of the foldr-build theorem, consider the following definitions.
\begin{verbatim}
   map f xs = build (\c n -> foldr (c . f) n xs)
   sum xs = foldr (+) 0 xs
   down m = build (\c n -> 
             let loop x = if x==0 then n else c x (loop (x-1))
             in loop m)
\end{verbatim}
where \verb@down n@ creates a list from \verb@n@ down to 1.
We could use these definitions to rewrite the expression \verb@sum (map sqr (down z))@ as follows
\begin{verbatim}
   sum (map sqr (down z))
     = foldr (+) 0 (build (\c n -> foldr (c . sqr) n (down z)))
     = foldr ((+) . sqr) 0 (down z)
     = let loop x = if x==0 then 0 else sqr x + loop (x-1)
       in loop z
\end{verbatim}
In just a couple of rewriting steps we have a purely recursive definition of the computation, with no intermediate lists. Obviously this is a simplistic example, but many examples of this form arise in practice, particularly when desugaring list comprehensions, or working with array expressions. And while programmers would not themselves write definitions in this form, all the primitive list-processing functions can be defined this way in the compiler's prelude, and many other definitions can be automatically transformed into this form \cite{LS95}.

\subsection{Inverses}\label{inverses}
The usual type for \verb@foldr@ is
\begin{verbatim}
   foldr :: (a -> b -> b) -> b -> [a] -> b
\end{verbatim}
If we reorder the arguments to place the list argument first, add explicit quantifiers for the type variables, and if we push the \verb@b@ quantifier in as much as possible, we get:
\begin{verbatim}
   foldr' :: forall a. [a] -> (forall b. (a -> b -> b) -> b -> b)
\end{verbatim}
Similarly, if we are explicit about the type of \verb@build@ we would write:
\begin{verbatim}
   build :: forall a. (forall b. (a -> b -> b) -> b -> b) -> [a]
\end{verbatim}
Now the foldr-build theorem reduces to simply stating that \verb@(foldr' . build) = id@. As it is also trivially the case that \verb@(build . foldr') = id@, we see that \verb@build@ and \verb@foldr'@ are inverses.

\subsection{Higher-order Folds}
The foldr-build technique works even for functions like \verb@reverse@ that are not expressible as first-order folds. First we define \verb@reverse@ as a higher-order fold,
\begin{verbatim}
   reverse xs = (foldr (\x k ys -> k (x:ys)) id xs) []
\end{verbatim}
in which the fold computation constructs a function which is applied to the extra argument \verb@[]@ supplied at the end, and then we abstract over the \verb@(:)@ and \verb@[]@ as follows,
\begin{verbatim}
   reverse xs = build (\c n -> (foldr (\x k p -> k (c x p)) id xs) n)
\end{verbatim}
In passing, we also $\alpha$-renamed \verb@ys@ to \verb@p@,
as the name \verb@ys@ suggests a list element, yet
in general the abstracted list structure is polymorphic and may not be building a list.
With this definition, list fusion proceeds exactly the same as before. For example, doing a variation to the previous derivation starting this time from the expression \verb@sum (reverse (map sqr (down z)))@ yields the following derivation\footnote{As before, this is a contrived example designed to show the \emph{form} of the code fusion transformation.}:
\begin{verbatim}
   sum (reverse (map sqr (down z)))
     = foldr (+) 0 (build (\c' n' -> (foldr (\x k p -> k (c' x p)) id 
                       (build (\c n -> foldr (c . sqr) n (down z))) 
                       n')))
     = foldr (\x k p -> k (x+p)) id (build (\c n -> foldr (c . sqr) n (down z))) 
                       0
     = foldr (\x k p -> k (sqr x + p)) id (down z) 0
     = (let loop x = if x==0 then (\p -> p) else (\p -> (loop (x-1)) (sqr x + p))
        in loop m) 0
     = let loop x p = if x==0 then p else loop (x-1) (sqr x + p)
       in loop m 0
\end{verbatim}
The result is a tight recursive definition with an accumulating parameter---we needed an $\eta$-expansion in the final step to obtain said parameter.

Note that, while \verb@foldr@ and \verb@foldl@ are mutually definable on flat structures such as arrays, they are not at all dual in the world of partial structures such as lazy lists.
In particular, just like \verb@reverse@, the \verb@foldl@ function is expressible 
exactly as a higher-order \verb@foldr@ as follows,
\begin{verbatim}
foldl :: (a -> b -> a) -> a -> [b] -> a
foldl g z xs = (foldr (\x k w -> k (g w x)) id xs) z
\end{verbatim}
The converse does not hold: \verb@foldr@ is \emph{not} expressible in terms of \verb@foldl@, as the result is always strict whereas \verb@foldr@ is not.

\subsection{Zip and Folds}
Foldr-build works beautifully for a huge range of list processing functions but comes to a crashing stop with \verb@zip@. Like with \verb@reverse@,
by using a higher-order instance of
\verb@foldr@ we can define zip as a fold on either one of the branches. The other list is passed as
an inherited attribute as follows:
\begin{verbatim}
   zip xs ys = build (\c n -> let c1 x g [] = n
                                  c1 x g (y:ys) = c (x,y) (g ys)
                              in
                                 foldr c1 (\ys -> n) xs ys)
\end{verbatim}
Unfortunately, using this technique leads us to define two asymmetric versions of zip. Using one
or the other of these we can fuse a left branch or a right branch computation, but
not both branches at the same time. It became accepted folklore that
zip cannot be defined as a fold on both branches at the same time. 

While this is true if we are restricted to ground terms or first-order functions, 
it is not true if we move to the the world of hyperfunctions.

\section{Coroutining Folds}

In order to handle accumulator functions like \verb@reverse@ or even \verb@foldl@, we had to write the functions as higher-order folds. That is, the \verb@foldr@ works in first-order function spaces such as
\verb@[a]->[a]@, constructing compositions of functions as it traverses its list argument. Interestingly, this is not as inefficient as it may sound if the final argument for the function is already present---as we saw in the derivation above, $\eta$-expansion flattens the function construction down to a use of an accumulating parameter.

Simple first-order constructions won't allow us to tame \verb@zip@, but with the power of universal domain equations behind us (e.g. $D \cong (D\rightarrow D)_\bot$) we have the flexibility to try much more ``interesting'' functional structures. In this spirit, we will depart from any obligation to satisfy any particular type system at this stage, and feel free to explore definitions that would be rejected by Haskell (so long as we can fix them up afterwards).

We will give \verb@fold@ an extra argument that behaves as a
coroutining continuation. This gives a (non-standard) definition of \verb@fold@ as follows:
\begin{verbatim}
   fold []     c n = \k -> n
   fold (x:xs) c n = \k -> c x (k (fold xs c n))
\end{verbatim}
Following the observation in Section \ref{inverses}, we provide the list argument first. The intuition behind the
definition is that the \verb@fold@ function receives an \emph{interleaving continuation {\tt k}},
applies the ``cons-function'' \verb@c@ to \verb@x@ and to the result of applying the continuation \verb@k@ to the recursive call of \verb@fold@. \emph{Note that, in the recursive call, the continuation {\tt k} is not provided.} Instead, \verb@k@ will accept the recursive call of \verb@fold@ as
its own interleaving continuation, and may subsequently call it with a new continuation. Thus
computations and continuations switch roles back and forth repeatedly.

To see this in practice, consider the case where the interleaving continuation is another
instance of \verb@fold@ itself.
\begin{verbatim}
   fold [1,2,3] c n (fold [7,8] d m)
     = c 1 (fold [7,8] d m (fold [2,3] c n))
     = c 1 (d 7 (c 2 (d 8 (c 3 (fold [] d m (...))))))
     = c 1 (d 7 (c 2 (d 8 (c 3 m))))
\end{verbatim}
The folds over the [1,2,3] and [7,8] elements each invoke the other in turn---like coroutines---and thereby produce the interleaving effect.

The \verb@fold@ example showed exactly two interleaving computations coroutining with one another. To generalize the idea to allow one, two, or more coroutining computations, we introduce the following two operations:
\begin{verbatim}
   self k = k self
   (f # g) k = f (g # k)
\end{verbatim}
These are both recursive definitions. The \verb@self@ function acts as a trivial continuation, and \verb@#@ acts as a composition operation which composes two coroutining continuations into a single coroutining continuation.

Notice, for example, that
the expression \verb@fold xs c n self@ is equal to \verb@foldr c n xs@. At each level of the
recursion, \verb@self@ simply hands control back to \verb@fold@ to operate on the next element, with itself as the next continuation to be invoked.

The composition operator \verb@#@ plays the dual role. If
we need to interleave three or more computations, we do so using \verb@#@. The combined
computation \verb@(f#g)@ when given a continuation \verb@k@ invokes \verb@f@ with continuation
\verb@(g#k)@.
When that continuation is invoked (assuming it ever is) then \verb@(g#k)@ will be applied
to some follow-on from \verb@f@, \verb@f'@ say. Then \verb@(g#k) f'@ will invoke 
\verb@g@ with continuation \verb@(k#f')@, and so on. So if we wanted to interleave three instances of \verb@fold@ we could do so as follows:
\begin{verbatim}
   fold [25] c n (fold [1,2,3] d m # fold [7,8] f p)
     = c 25 ((fold [1,2,3] d m # fold [7,8] f p) (fold [] c n))
     = c 25 (fold [1,2,3] d m (fold [7,8] f p # fold [] c n))
     = c 25 (d 1 ((fold [7,8] f p # fold [] c n) (fold [2,3] d m)))
     = c 25 (d 1 (fold [7,8] f p (fold [] c n # fold [2,3] d m)))
     = c 25 (d 1 (f 7 ((fold [] c n # fold [2,3] d m) (fold [7,8] f p))))
     = c 25 (d 1 (f 7 (fold [] c n (fold [2,3] d m) (fold [7,8] f p))))
     = c 25 (d 1 (f 7 n))
\end{verbatim}
Naturally \verb@self@ and \verb@#@ behave well together, as expressed in the following theorem:
\begin{theorem}
\verb@#@ is associative, and \verb@self@ is a left and right identity for \verb@#@.
\end{theorem}
Using this result, the interleaving of folds above can be written as
\begin{verbatim}
   (fold [25] c n # fold [1,2,3] d m # fold [7,8] f p) self
\end{verbatim}
The power of this version of \verb@fold@ is apparent when we see that \verb@zip@ is
definable in terms of independent \verb@fold@s on its two branches, thus:
\begin{verbatim}
   zip xs ys = (fold xs first []  #  fold ys second Nothing) self
     where
       first x Nothing        = []
       first x (Just (y,xys)) = (x,y) : xys
       second y xys = Just (y,xys)
\end{verbatim}
Note that while \verb@first@ is strict---the presence or absence of a \verb@y@ element is needed
to know whether to produce a new output element---the \verb@second@ function is not, thus
\verb@zip@ has its usual non-strict behavior.

\subsection{Universal Domains}
What kind of functions are \verb@self@ and \verb@#@? The definition of \verb@self@ requires a solution to the function equation $t_1 = (t_1 \rightarrow t_0) \rightarrow t_0$, and a similar equation arises from within the definition of \verb@#@. Moreover, the definition of \verb@fold@ also requires an infinite type. In this case, the general form of the type equation that needs to
be solved is \verb@H a b = H b a -> b@, which when expanded gives the infinite type
\begin{verbatim}
   H a b = ((((... -> a) -> b) -> a) -> b
\end{verbatim}
Clearly no set-theoretic solution exists for these, but they live quite happily within any domain that contains an image of its own function space, such as $D$ above. It is instructive to consider to CPO-theoretic approximations to H. The first is
\begin{verbatim}
   H0 a b = 1
   H1 a b = (1 -> a) -> b
          = a -> b
   H2 a b = (H1 a b -> a) -> b
          = ((a -> b) -> a) -> b
\end{verbatim}
Interestingly, all the bs are in positive (covariant) positions and all the as are
in negative (contravariant) positions. In effect, the a's act as arguments, and
the b's act as results. That is, the type \verb@H a b@ is a kind of function from a to
b, or even a stack of functions from a to b. We use the term \emph{hyperfunction}
to express this.

\section{Typing}

To code hyperfunctions in Haskell we introduce \verb@H@ as a newtype, and define
appropriate access functions.
\begin{verbatim}
   newtype H a b = Fn {invoke :: H b a -> b}

   (#) :: H b c -> H a b -> H a c
   f # g = Fn (\k -> invoke f (g # k))

   self :: H a a                        
   self = lift id

   lift :: (a->b) -> H a b
   lift f = f << lift f

   (<<) :: (a -> b) -> H a b -> H a b
   f << q = Fn (\k -> f (invoke k q))

   base :: a -> H b a
   base p = Fn (\k -> p)

   run :: H a a -> a
   run f = invoke f self
\end{verbatim}
The use of the explicit constructor \verb@Fn@ and deconstructor \verb@invoke@ obscure some of the definitions a little, though they provides us with informative types in return. In particular, the type of \verb@#@ makes it clear that it is acting as a composition operator.  In fact, hyperfunctions form a category over the same objects as the base functions use, and \verb@lift@ is a functor from the base category into the hyperfunction
category. The \verb@lift@ operator takes a normal function \verb@f@ and turns it into a hyperfunction
by acting as \verb@f@ whenever it is invoked. If we were to expand its definition (and
again present it untyped) we get \verb@lift f k = f (k (lift f))@. Interestingly the \verb@self@ operator is simply an instance of \verb@lift@.

We defined \verb@lift@ using a new operator \verb@(<<)@, which
acts rather like a cons operator by taking a function element \verb@f@ and adding to the
``stack'' of functions \verb@q@ (and invoking an intervening coroutining continuation in-between, of course). Without types we could define \verb@(<<)@ by \verb@(f<<q) k = f (k q)@.

These definitions make it easy to define a hyperfunction form of the fold operation:
\begin{verbatim}
   fold :: [a] -> (a -> b -> c) -> c -> H b c
   fold [] c n     = base n
   fold (x:xs) c n = c x << fold xs c n
\end{verbatim}
The (re-)definition of \verb@fold@ makes it clear that it is an instance of the usual
\verb@foldr@ as follows. In fact, the following two equations hold
\begin{verbatim}
   fold xs c n = foldr (\x z -> c x << z) (base n) xs
   foldr c n xs = run (fold xs c n)
\end{verbatim}
This ability to define either in terms of the other shows that \verb@fold@ and \verb@foldr@
are equivalent to one another.

\section{Fold-Build}

As in the original foldr-build work, we now define a build function which ensures
that the list generator function is suitably abstracted.
\begin{verbatim}
   build :: (forall (b,c).(a->b->c) -> c -> H b c) -> [a]
   build g = run (g (:) [])
\end{verbatim}
Our conjecture is that under appropriate circumstances, the fusion law holds:
\begin{verbatim}
   fold . build = id
\end{verbatim}
Initially we hoped that the ``appropriate circumstances'' would simply be the
parametric nature of \verb@g@'s type. Now it appears we'll have to be slightly more
clever. For example, restricting \verb@g@ to be constructed by repeated applications of
\verb@<<@ and \verb@base@ turns out to be sufficient, and this could be achieved by making
the \verb@H@ type abstract. In this paper, the fusion law is left as a conjecture requiring further characterization, and we will focus instead on better characterizing the hyperfuntions
themselves.
Just before we do, let us see the fusion law in practice---now with zip.
Consider the example of fusing 
\begin{verbatim}
   sum (zipW (*) (map sqr xs) (map inc ys))
\end{verbatim}
where \verb@zipW@ is a \verb@zipWith@-like function whose definition is similar to the
definition of \verb@zip@ we saw earlier:
\begin{verbatim}
   zipW f xs ys = build (zipW' f xs ys)
   zipW' f xs ys c n = fold xs first n  #  fold ys second Nothing
     where
       first x Nothing        = n
       first x (Just (y,xys)) = c (f x y) xys
       second y xys = Just (y,xys)
\end{verbatim}
In this case, the zipper-function \verb@f@ is applied within the right-hand side of \verb@first@ whenever a pair of entries from the two lists are present. The other list processing functions are defined as follows:
\begin{verbatim}
   map f xs = build (\c n -> fold xs (c . f) n)
   sum xs = run (fold xs (+) 0)
\end{verbatim}
The fusion proceeds through beta reduction, symbolic composition, and application of fold-build.
\begin{verbatim}
   sum (zipW (*) (map sqr xs) (map inc ys))
   = run (fold (zipW (*) (map sqr xs) (map inc ys)) (+) 0)
   = run (fold (map sqr xs) first 0 # fold (map inc ys) second Nothing)
       where
         first x Nothing        = 0
         first x (Just (y,xys)) = (x * y) + xys
         second y xys = Just (y,xys)
   = run (fold xs (first . sqr) 0  # fold ys (second . inc) Nothing)
       where
         first x Nothing        = 0
         first x (Just (y,xys)) = (x * y) + xys
         second y xys = Just (y,xys)
   = run (fold xs first' 0 # fold ys second' Nothing)
       where
         first' x Nothing        = 0
         first' x (Just (y,xys)) = (sqr x * y) + xys
         second' y xys = Just (inc y, xys)
\end{verbatim}
The intermediate lists produced by the two uses of \verb@map@ have both been fused
away, even though they occurred in separate branches of the \verb@zip@.

A recap on the context of this approach is probably useful at this
point. Simple folds at ground types (that is, that build non-function structures) have been well studied. These uses of fold
construct synthesized attributes only. Once we allow folds to build functions as their results,
we gain significant extra power. The function arguments allow us to model
inherited attributes that allow \verb@reverse@ and even \verb@foldl@ to be seen as instances of \verb@foldr@. Once we allow folds to build hyperfunctions, we enable \emph{coroutining} between distinct fold computations. This appears to go beyond the usual language of
attribute grammars with inherited and synthesized attributes as it now permits
attributes to flow between the nodes of different trees that are at the same
level, and a (nearly) symmetric definition of \verb@zip@ becomes possible. 

\section{Hyperfunctions Axiomatically}
The operators we defined on hyperfunctions encourage us to move away from the explicit model-theoretic view of H, and consider instead an axiomatic approach. This will allow us to consider other models which may be more efficient implementations in certain cases.

We continue to use the notation \verb@H a b@, but now for an abstract type of hyperfunctions. We regard \verb@H a b@ as describing the set (or more likely, the CPO) of arrows between objects \verb@a@ and \verb@b@ in an new \emph{hyperfunction} category which shares the same objects as the original. We require the following operations:
\begin{verbatim}
   (#)  :: H b c -> H a b -> H a c
   lift :: (a->b) -> H a b
   run  :: H a a -> a
   (<<) :: (a->b) -> H a b -> H a b
\end{verbatim}
which must satisfy the following conditions
\begin{verbatim}
   Axiom 1.  (f # g) # h = f # (g # h)
   Axiom 2.  f # self = f = self # f
   Axiom 3.  lift (f . g) = lift f # lift g
   Axiom 4.  run (lift f) = fix f

   Axion 5.  (f << p) # (g << q) = (f . g) << (p # q)
   Axiom 6.  lift f = f << lift f
   Axiom 7.  run ((f << p) # q) = f (run (q # p))
\end{verbatim}
where \verb@self :: H a a@ is defined by \verb@self = lift id@, and \verb@fix@ and composition have their usual meanings. These axioms make hyperfunctions into a category. The \verb@lift@ function is a functor from the base category into the hyperfunction category, and \verb@(lift f)@ lets us see an underlying function \verb@f@ as a hyperfunction. The \verb@#@ operation extends the composition, and \verb@run@ extends the fix point operator. The following definition of \verb@mapH@
\begin{verbatim}
   mapH :: (a' -> a) -> (b -> b') -> H a b -> H a' b'
   mapH r s f = lift s # f # lift r   
\end{verbatim}
demonstrates that \verb@H@ is contravariantly functorial in its first argument, and covariantly in its second.

Using these definitions, we can now define some of the other hyperfunction operations:
\begin{verbatim}
   invoke :: H a b -> H b a -> b
   invoke f g = run (f # g)

   base :: b -> H a b
   base k = lift (const k)
\end{verbatim}
where \verb@const k = \x -> k@ is the constant function.
It now follows that \verb@run f = invoke f self@, giving a definition of \verb@run@ whenever \verb@invoke@ is more naturally defined as a primitive (as in the earlier definition).

Without \verb@<<@ and its axioms, the system has a trivial model in which \verb@H a b = a -> b@, with \verb@#@ being ordinary function composition, \verb@lift f = f@, and so on. With \verb@<<@, the trivial model is no longer possible.

All hyperfunction models have the property that distinct functions remain distinct when regarded as hyperfunctions. Any category of hyperfunctions thus contains a faithful copy of the base category of ordinary functions.
\begin{theorem} 
The functor \verb@lift@ is faithful (i.e. if \verb@lift f = lift g@ then \verb@f = g@).
\end{theorem}
The theorem follows by an easy calculational proof. We define:
\begin{verbatim}
   project :: H a b -> (a->b)
   project q x = invoke q (base x)
\end{verbatim}
It suffices to show that \verb@project@ is a left-inverse of \verb@lift@, i.e. that \verb@project (lift f) = f@. Indeed:
\begin{verbatim}
   project (lift f) x 
    = invoke (lift f) (base x)
    = run (lift f # base x)
    = run ((f << lift f) # base x)
    = f (run (base x # lift f))
    = f (run (lift (const x) # lift f)) 
    = f (run ((const x << base x) # lift f))
    = f (const x (run (lift f # base x)))
    = f x
\end{verbatim}
as required.

\section{A Stream Model for H}

Now that we view H as an abstract type, we are free to investigate alternative models
for it. We have two other models which provide useful insights into the core
functionality required for \verb@zip@ fusion.

The elements of H we have been using behave like a stream of functions: some initial portion of work
is performed, and then the remaining work is delayed and given to the continuation to be invoked at some point in the future (if at all). When
the continuation reinvokes the remainder, a little more work is done and again the
rest is given to \emph{its} continuation. In other words, work is performed piece by
piece with interruptions allowing for interleaved computation to proceed.

This intuition leads us to represent this family of hyperfunctions explicitly as an infinite
stream. We use the name \verb@L@ for this model.
\begin{verbatim}
   data L a b = (a->b) :<<: L a b

   invoke :: L a b -> L b a -> b
   invoke fs gs = run (fs # gs)

   (#) :: L b c -> L a b -> L a c        
   (f :<<: fs) # (g :<<: gs) = (f . g) :<<: (fs # gs)           

   self :: L a a
   self = lift id

   lift :: (a->b) -> L a b               
   lift f = f :<<: lift f

   base :: a -> L b a
   base x = lift (const x)

   (<<) :: (a->b) -> L a b -> L a b                    
   (<<) = (:<<:)                            

   run :: L a a -> a
   run (f :<<: fs) = f (run fs)
\end{verbatim}

One interesting aspect of this model is that \verb@run@ is more naturally primitive
than \verb@invoke@, whereas in the original function-space model \verb@H@
the opposite was the case. On the
other hand, the identity and associativity laws between \verb@#@ and \verb@self@
become very easy to prove just by fixed
point induction and properties of composition. In contrast, the corresponding
theorems about the H model turned out to be rather challenging, to say the least
\cite{KLP01}.

The stream of functions acts like a fixpoint waiting to happen. Two things could
occur: either the functions are interspersed with another stream of functions,
or all the functions are composed together by \verb@run@. In this way, \verb@run@
ties the recursive knot, and removes opportunities for further coroutining.

The \verb@fold@ function is defined exactly as before in terms of \verb@<<@ and
\verb@base@. Its behavior is given as follows:
\begin{verbatim}
   fold [x1,x2,x3] c n 
     = c x1 :<<: c x2 :<<: c x3 :<<: const n :<<: ...
\end{verbatim}
where the \verb@...@ indicates an infinite stream of \verb@const n@. Thus \verb@fold@
turns a list of elements into an infinite stream of partial applications of the
\verb@c@ function to the elements of the list. At this point we might ask whether
we have actually gained anything. After all, we have simply converted a
list into a stream. Even worse, the much vaunted definition of \verb@zip@ turns
out to be defined in terms of \verb@#@, which is defined just like \verb@zip@ in
the first place! However, the stream is merely intended to act as a temporary
structure which helps the compiler perform its optimizations. As with the original
type H of hyperfunctions, the
L model can be also used for fold-build fusion, and the stream structures are
optimized away. Any that exist after the fusion phase may (in principle) be removable
by inlining the definition of \verb@run@. In other words, if the compiler is able to clean
up sufficiently, the stream structure
simply will not exist at run-time---it's purpose is compile-time only.

Though we won't prove it here, the L model is the simplest possible model of hyperfunctions. Formally, in the category of hyper function models, the L model is an initial object. The L model seems to capture something essential about using hyperfunctions to define \verb@zip@. It expresses
the ``linear'' behavior of \verb@fold@ as it traverses its input lists.

\section{A State-Machine Model for Hyperfunctions}

We need to go a little further than L to find a good model for doing zip-fusion in practice. One strength of the original foldr-build is that it could fuse with recursive
generators for lists, often ending up with computations that had no occurrence of
lists whatsoever. In one sense it was easy. By restricting to fusion along a
single branch of zip-like functions, we always ended up with a single ultimate recursive
origin for the computation. All the foldr-build rules had to do was place in the subsequent processing of list elements into the appropriate places within this (arbitrarily recursive) structure, and we were done.

In contrast, multi-branch fusion may have many sources each acting as a partial origin of the computation, so we may need to to combine multiple recursive
generators. This is very hard in general, so to make the problem tractable we
focus on recursive generators that are state machines, also known as tail calls or
anamorphisms. This leads us to yet another model for hyperfunctions where we represent the
state of a function as an anamorphism. The type of this state element can be
hidden by using a rank-2 universally quantified type.
\begin{verbatim}
   data A a b where
      Hide :: (u -> Either b (a -> b,u)) -> u -> A a b

   lift :: (a->b) -> A a b
   lift f = Hide (\u -> Right (f, u)) (error "Null")

   (#) :: A b c -> A a b -> A a c
   Hide g x  #  Hide g' x'
     = Hide (\(z,z') -> case g z of
                 Left n     -> Left n
                 Right(f,y) -> case g' z' of
                      Left m        -> Left (f m)
                      Right(f',y')  -> Right (f . f', (y,y')))
            (x,x')

   run :: A a a -> a                      
   run (Hide f v) = loop v     
     where
       loop x = case f x of
                  Left n     -> n
                  Right(h,y) -> h (loop y)

   (<<) :: (a->b) -> A a b -> A a b
   p << (Hide f v)
     = Hide (\x -> case x of
                     Nothing -> Right (p, Just v)
                     Just w  -> case f w of
                                  Left n      -> Left n
                                  Right (h,y) -> Right (h, Just y))
            Nothing
\end{verbatim}
The form of the type declaration uses the GADT syntax. The declaration gives \verb@Hide@ a type in which \verb@u@ is universally quantified. By the usual interchange laws between universal and existential quantifiers, this is equivalent to
\begin{verbatim}
   Hide :: (exists u . (u -> Either b (a -> b,u)), u) -> A a b
\end{verbatim}
In other words, the objects of the A model are state machines whose inner state is completely hidden from the outside world, except inasmuch as they produce the next portion of a coroutining function on demand.

Going back to zip-fusion, we can put these definitions to work. We define a couple of typical generators.
\begin{verbatim}
   down z = build (down' z)

   down' :: Int -> (Int -> b -> c) -> c -> A b c
   down' w c n = Hide (\z -> if z<=0 then Left n
                             else Right (c z,z-1))
                      w

   upto i j = build (upto' i j)

   upto' :: Int -> Int -> (Int -> b -> c) -> c -> A b c
   upto' a b c n = Hide (\(i,j) -> if i>j then Left n
                                   else Right (c i,(i+1,j)))
                        (a,b)
\end{verbatim}
As an example, we fuse the expression \verb@sum (zipW (*) (upto 2 10) (down 6))@. The derivation 
proceeds as follows.
\begin{verbatim}
   sum (zipW (*) (upto 2 10) (down 6))
    = run (fold (zipW (*) (upto 2 10) (down 6)) (+) 0)
    = run (zipW' (*) (upto 2 10) (down 6) (+) 0)
    = run (fold (upto 2 10) c 0  #  fold (down 6) d Nothing)
      where
        c x Nothing        = 0
        c x (Just (y,xys)) = (x * y) + xys
        d y xys = Just (y,xys)
    = run (upto' 2 10 c 0  #  down' 6 d Nothing)
      where
        c x Nothing        = 0
        c x (Just (y,xys)) = (x * y) + xys
        d y xys = Just (y,xys)
    = run (Hide (\(i,j) -> if i>j then stop 0 else (c i,(i+1,j)))
                (2,10)
           #
           Hide (\z -> if z<=0 then stop Nothing else (d z,z-1))
                6)
      where
        c x Nothing        = 0
        c x (Just (y,xys)) = (x * y) + xys
        d y xys = Just (y,xys)
    = run (Hide (\((i,j),z) -> case if i>j then Left 0 else Right (c i,(i+1,j)) of
                                  Left n      -> Left n
                                  Right (f,y) ->
                               case if z<=0 then Left Nothing else Right (d z,z-1) of
                                  Left m        -> Left (f m)
                                  Right (f',y') -> (f . f', (y,y')))
                  ((2,10),6))
      where
        c x Nothing        = 0
        c x (Just (y,xys)) = (x * y) + xys
        d y xys = Just (y,xys)
   = run (Hide (\((i,j),z) -> if i>j then Left 0 else
                              let (f,y) = (c i,(i+1,j)) in
                              if z<=0 then Left (f Nothing) else
                              let (f',y') = (d z,z-1) in
                              (f . f', (y,y')))
                ((2,10),6))
     where
       c x Nothing        = 0
       c x (Just (y,xys)) = (x * y) + xys
       d y xys = Just (y,xys)
    = run (Hide (\((i,j),z) -> if i>j then stop 0 else
                               if z<=0 then stop 0 else
                               (\w -> (i*z)+w, ((i+1,j),z-1)))
          ((2,10),6))
    = loop ((2,10),6)
      where
        loop ((i,j),z) = if i>j then 0 else
                         if z<=0 then 0 else
                         (i*z) + loop ((i+1,j),z-1)
\end{verbatim}
The first few steps are just the fold-build from before, and the remainders normal program simplification. The result is quite impressive. All intermediate lists have been removed, and the multiple generators merged into a single generator.

\section{Conclusion}

The original motivation for hyperfunctions was to broaden the power of the fold-build
fusion technique to be able to handle multiple input lists. In this it succeed, and
we have demonstrated  that many occurrences of \verb@zip@ can be eliminated using the fold-build technique, leading to the fusion of multiple list generation routines. The implementation is simple
and it works well, but as yet it is not clear how well it would work in large
examples. Of course, as noted in the introduction,
the whole approach of foldr-build has been eclipsed by the stream fusion techniques \cite{CLS07}. In stream fusion, lists are represented as (non-recursive) state-machine stream processors, and it turns out to be quite feasible to fuse these state-machines together, including for the case of zip.

However, coroutining folds may turn out to have merit in their own right. In particular,
the real insights of this papers are twofold:

First, we were forced to realize
that the \verb@fold@ function is even more powerful than we had previously thought.
In particular, it came as a palpable shock that \verb@fold@ was able to express
interleaving computations. The view of \verb@fold@ as a generic expression simply of
inherited and synthesized attributes over tree shaped structures had become quite
deeply ingrained. Whether this understanding of the coroutining capability of
\verb@fold@ will lead to new functions and techniques remains to be seen, but it
cannot but help in broadening our perspectives.

Secondly, even though we have moved to use other models as well, we have found
hyperfunctions fascinating in their own right. They have been devilishly tricky
to reason about directly, but now we know that they form a category, have a weak
product and seem to fit nicely into Hughes arrow class \cite{Hug00}. Again,
whether they will turn out to be useful in other applications remains to be seen.

\section{Acknowledgements}

An early version of these ideas was presented to the IFIP working group on functional
programming (WG2.8, 1999) where we received helpful feedback from many people, but
especially Simon Peyton Jones, Richard Bird, Ralph Hinze, Erik Meijer.

\bibliographystyle{eptcs}

\providecommand{\urlalt}[2]{\href{#1}{#2}}
\providecommand{\doi}[1]{doi:\urlalt{http://dx.doi.org/#1}{#1}}

\end{document}